# Towards multigrid methods for propagators of staggered fermions with improved averaging and interpolation operators

T. Kalkreuter[a]*

[a]II. Institut für Theoretische Physik, Universität Hamburg,
Luruper Chaussee 149, D-22761 Hamburg, Germany

A Dirac choice for the averaging kernel $C$ is implemented numerically. This improved kernel will be needed in gauge covariant multigrid computations for propagators of staggered fermions. Results for $C$ and the variational coarse grid operator will be given in 2-$d$ $SU(2)$ gauge fields. C++ is advocated for future algorithm development.

Big efforts have been undertaken to find efficient multigrid (MG) methods for the computation of propagators in gauge fields; see [1] for a list of references. Up to now no practical method has been found which is competitive for fermionic propagators in 4-$d$ non-abelian gauge fields. However, it was proven that (geometric) MG works in principle without critical slowing down (CSD) in arbitrarily disordered systems, both for bosons [2,3] and for staggered fermions [4]. Unfortunately, this "ideal" algorithm is not practical because the interpolation kernels have exponential tails whose truncation brings CSD back [3]. Nevertheless, the proof that CSD can be eliminated by MG raises hope that there exist also practical – though sophisticated – competitive methods.

We consider the squared Dirac equation,

$$(-\slashed{D}^2 + m^2)\chi = f \ , \qquad (1)$$

where $f$ may be a pseudofermion field, for instance. We will focus on staggered fermions and a gauge covariant "Dirac choice" [5] (no gauge fixing required) for the averaging kernel $C$. The $C$-kernel should map smooth functions on a fine grid onto functions which are smooth on a coarser scale. In [6,7] it was proposed to adopt an algebraic definition of smoothness in disordered systems.

Given a differential operator $D$, according to [6,7] a function $\chi$ is smooth on length scale $a$ when

$$\|D\chi\| \ll \|\chi\| \ \ (\text{in units } a = 1) \ . \qquad (2)$$

*Work supported by Deutsche Forschungsgemeinschaft.
To appear in the proceedings of LATTICE '93.

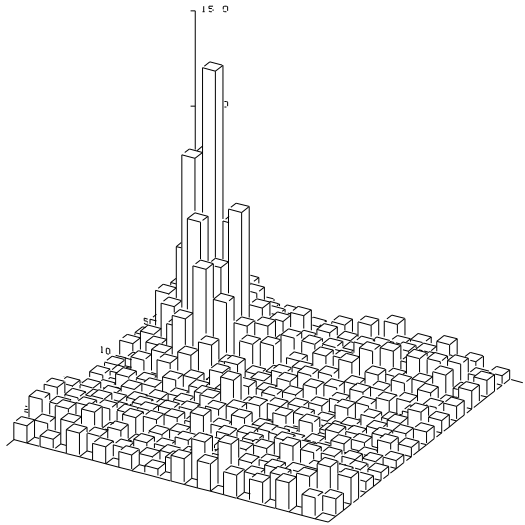

Figure 1. Norm of the ground-state of $-\slashed{D}^2$ on the even sublattice of a $24^4$ lattice in a quenched $SU(2)$ gauge field at $\beta = 2.5$. (Two-dimensional cut through the site with maximal norm.)

*Smooth functions in the sense of Eq. (2) are localized in disordered systems.* Bäker [8] found in quenched 2-$d$ $SU(2)$ lattice gauge theory that the eigenfunctions corresponding to the 2 or 3 lowest eigenvalues of the negative gauge covariant Laplacian are localized for $\beta = O(1)$. A fermionic example in 4-$d$ $SU(2)$ is given in Fig. 1.

The localization property of the lowest mode of $-\slashed{D}^2$ could be taken into account in a geometric MG algorithm by shifting the site with maximal norm such that it coincides with a block center.

Let us now turn to the improved averaging kernel $C$. In Ref. [5] two qualitatively different proposals were made for the choice of $C$ for staggered fermions in non-abelian gauge fields. The first one, the "Laplace choice" (LC), was used successfully in the "ideal" algorithm [4], but it has proven to be poor in practical variational MG methods [1,3]. Therefore we will test the second proposal which is called the "Dirac choice" (DC) of $C$.

In the DC, as in the LC, $C$ is defined by a ground-state projection method. This means that the adjoint $C^*$ of $C$ fulfills a gauge covariant eigenvalue equation,[1]

$$(-\slashed{D}^2_{N,D,x} C^*)(z,x) = \lambda_0(x) C^*(z,x) , \qquad (3)$$

where $-\slashed{D}^2_{N,D,x}$ is a block-local approximation of $-\slashed{D}^2$ specified below, and $\lambda_0(x)$ is its lowest (gauge invariant) eigenvalue. $z$ denotes a site in a fine lattice, and $x$ is a coarse grid site ("a block"). $C(x,z)$ is an element of the linear span of the gauge group.

The solution of Eq. (3) is made unique by imposing a normalization and a covariance condition. The normalization condition reads[2]

$$\frac{1}{L_b^d} \sum_{z \in x} \|C(x,z)\|^2 = 1 \qquad (4)$$

for all $x$, where $L_b$ is the blocking factor ($= 3$), summation is over all fine grid sites within a block $x$, and the norm used is the trace norm. The covariance condition reads (for gauge group $U(1)$ or $SU(2)$)

$$C(x,\hat{x}) = r(x) \mathbb{1} \qquad (5)$$

with $\hat{x}$ being the block center, and $r(x) > 0$.

We use a blocking procedure for staggered fermions which is consistent with the lattice symmetries of free fermions [5]. This forces us to choose $L_b = 3$. Even $L_b$ are not allowed.

Fig. 2 illustrates our choice of blocks in case of the DC. The different fermionic degrees of freedom are called "pseudoflavor" [5]. Different pseudoflavors are distinguished by different symbols in Fig. 2. Block centers $\hat{x}$ are encircled. They are spaced by $L_b$, which means that the coarser grid has $1/L_b^d$ times the number of sites of the finer grid. The boundaries of seven blocks $x$ are marked.

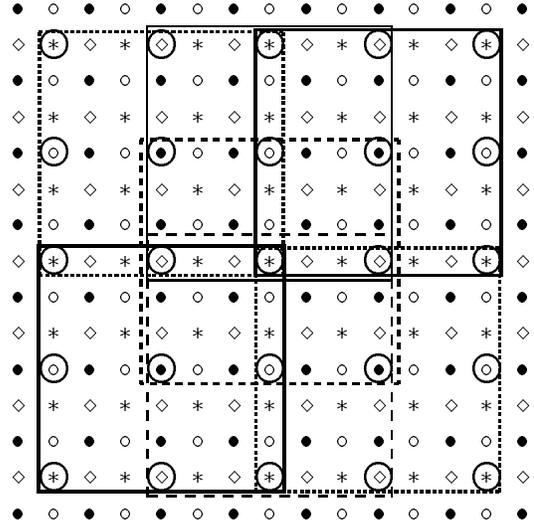

Figure 2. Dirac-choice-blocking of a two-dimensional staggered lattice.

The operator $\slashed{D}^2_{N,D,x}$ in (3) equals $\slashed{D}^2$ with the following boundary conditions (BC). For the sites which have the same pseudoflavor as the central site $\hat{x}$ we impose Neumann BC on the boundary of $x$, for all other boundary sites we use Dirichlet BC. Details can be found in [5,3].

The DC (3) has the following properties. In the limiting case of a pure gauge, we regain the LC with nonoverlapping blocks. The $C$-kernel for free fermions is piecewise constant on sublattices of given pseudoflavor.[3] In nontrivial gauge fields – contrarily to the LC – the averaging kernel $C(x,z)$ of the DC is nonvanishing also when $x$ and $z$ carry different pseudoflavor (but have the same parity, i. e. are both even or both odd). Therefore the blocks in Fig. 2 overlap in a nontrivial way. This overlap takes the field strength term $\sigma_{\mu\nu}F_{\mu\nu}$ in $\not{D}^2$ into account. $\sigma_{\mu\nu}F_{\mu\nu}$ had been disregarded in the LC.

---

[1] $C$ depends on the gauge field although this is not indicated explicitly.

[2] Because of the nontrivial overlap of blocks in the DC, we cannot retain the previous condition $CC^* = \mathbb{1}$. However, in pure gauges nothing changes.

[3] In pure gauges the variational MG algorithm with these kernels is successful in eliminating CSD [3].

The DC for $C$ has been implemented in C++ [9,10]. The author advocates this language for future algorithm development. It requires some effort to define suitable C++ classes, but once this is done, all advantages of C++ [10, Sec. 1] are at your disposal.

The eigenvalue equation (3) was solved by the efficient algorithm of [11]. All computer programs were checked for gauge covariance (i. e. without gauge fixing corresponding results were found for different points on gauge orbits).

First tests have been performed in quenched 2-$d$ $SU(2)$ gauge fields. An encouraging result is that $\lambda_0(x) = 0$ *for all blocks $x$ and for arbitrary $\beta$*. Global lowest eigenvalues are collected in Table 1. Examples for $C$, for the variational coarse grid operator $-C\not{D}^2C^*$, and for the non-diagonal(!) mass term $CC^*$ are in Table 2. Further results in $d = 2, 4$, including computations of propagators, will be available shortly [12].

I am indebted to G. Mack for many stimulating discussions. Financial support by Deutsche Forschungsgemeinschaft is gratefully acknowledged. I wish to thank DESY for providing resources on HP workstations .

Table 1
Lowest eigenvalues; cf. Table 1 of [4].

| $\beta$ | $|\Lambda|$ | $-\not{D}^2$ | $-C\not{D}^2C^*$ |
|---|---|---|---|
| $\infty$ | any | 0 | 0 |
| 40 | $24^2$ | 0.004513 | 0.206828 |
| 1.0 | $24^2$ | 0.000173 | 0.418796 |

Table 2
Examples of kernels in 2-$d$ $SU(2)$ at $\beta = 40$ ($\xi \approx 5.2$). The bold entries vanish in pure gauges.

| $\|C(0,z)\|$ | | | | |
|---|---|---|---|---|
| 0.997 | 0 | 0.996 | 0 | 0.993 |
| 0 | **0.052** | 0 | **0.203** | 0 |
| 1.00 | 0 | 1.01 | 0 | 1.10 |
| 0 | **0.129** | 0 | **0.164** | 0 |
| 0.981 | 0 | 0.984 | 0 | 0.987 |
| $3\|-C\not{D}^2C^*(0,x)\|$ | | | | |
| 0 | 0 | 0.910 | 0 | 0 |
| 0 | **0.266** | 0 | **0.226** | 0 |
| 0.904 | 0 | 3.94 | 0 | 0.941 |
| 0 | **0.179** | 0 | **0.146** | 0 |
| 0 | 0 | 0.913 | 0 | 0 |
| $3\|CC^*(0,x)\|$ | | | | |
| | **0.060** | 0 | **0.130** | |
| | 0 | 3.00 | 0 | |
| | **0.069** | 0 | **0.078** | |

### REFERENCES

1. T. Kalkreuter, Nucl. Phys. B (Proc. Suppl.) 30 (1993) 257.
2. T. Kalkreuter, Phys. Lett. B276 (1992) 485.
3. T. Kalkreuter, *Ph.D. thesis* and preprint DESY 92–158.
4. T. Kalkreuter, Phys. Rev. D48 (1993) R1926.
5. T. Kalkreuter, G. Mack, and M. Speh, Int. J. Mod. Phys. C3 (1992) 121.
6. M. Bäker, T. Kalkreuter, G. Mack, and M. Speh, in: Proceedings of the 4th International Conference on Physics Computing PC '92, eds. R.A. de Groot and J. Nadrchal (World Scientific, Singapore, 1993).
7. M. Bäker, G. Mack, and M. Speh, Nucl. Phys. B (Proc. Suppl.) 30 (1993) 269.
8. M. Bäker, in preparation.
9. B. Stroustrup, *The C++ programming language* (Addison-Wesley, Reading, 1992).
10. B. Eckel, *C++ Inside & Out* (McGraw-Hill, Berkeley, 1993).
11. T. Kalkreuter, Nucl. Phys. B376 (1992) 637.
12. T. Kalkreuter, in preparation.